\documentclass[notitlepage]{revtex4-1}

\usepackage[utf8]{inputenc}
\usepackage[colorlinks=true,citecolor=blue,linkcolor=blue]{hyperref}
\usepackage[normalem]{ulem}
\usepackage{url}
\usepackage{graphicx,wrapfig,float,slashed,cancel}
\usepackage{amsmath,amssymb,epsfig,graphicx,xcolor,stmaryrd}
\usepackage{bm}
\usepackage{enumitem}
\usepackage{orcidlink}
\begin{document}

\title{Doubly heavy spin-$\frac {3}{2} $ baryons spectrum in the ground and excited states
}

	\author{M.~Shekari Tousi$^{a}$\orcidlink{0009-0007-7195-0838}}
    \email{marzieshekary@gmail.com}
	
	\author{K.~Azizi$^{a,b}$\orcidlink{0000-0003-3741-2167}} 
	 \email{kazem.azizi@ut.ac.ir} \thanks{Corresponding author} 
	
	\affiliation{
		$^{a}$Department of Physics, University of Tehran, North Karegar Avenue, Tehran 14395-547, Iran\\
		$^{b}$Department of Physics, Dogus University, Dudullu-\"{U}mraniye, 34775
		Istanbul, T\"{u}rkiye
	}
	
\date{\today}

\preprint{}

\begin{abstract}
	
	This study employs QCD sum rules to predict the masses and residues of spin-$\frac {3}{2} $ doubly heavy baryons including two heavy quarks (c and/or b) and one light quark, specifically focusing on $	\Xi_{cc}^*$, $	\Xi_{bc}^*$, $	\Xi_{bb}^*$, $	\Omega_{cc}^*$, $	\Omega_{bc}^*$ and $	\Omega_{bb}^*$. Our study provides results for the ground state ($1S$), first orbital excitation ($1P$), and the first radial excitation ($2S$), within a consistent theoretical framework. In addition to mass spectra, we provide residue calculations  as well. The calculated residues are essential for estimating the decay widths and branching ratios of these baryons at different decay   channels. Our analysis incorporates nonperturbative QCD effects through operators up to dimension ten, leading to improved precision in the mass and residue calculations. These predictions offer crucial guidance for ongoing and future experimental searches, particularly in light of the current lack of empirical data for  the ground and excited states, and provide a basis for comparison with future experimental data.

\end{abstract}


\maketitle

\renewcommand{\thefootnote}{\#\arabic{footnote}}
\setcounter{footnote}{0}
\section{\label{sec:level1}Introduction}\label{intro} 

The existence of baryons composed of two heavy quarks has presented a long-standing challenge in the quark model ~\cite{GellMann:1964nj}, as experimental observations have been remained elusive. This observational gap was addressed in 2002 with the SELEX Collaboration’s identification of the $\Xi^+_{cc}(3520)$ through its $ p D^+ K^-$  decay  \cite{SELEX:2002wqn}. Their findings were corroborated by the same group in 2005 \cite{SELEX:2004lln}. A significant advancement came  when the LHCb reported the exploration of $\Xi^{++}_{cc}(3621)$ in 2017, identified via  $ \Xi^{++}_{cc}\rightarrow \Lambda^+_{c} K^- \pi^+ \pi^+$ decay \cite{LHCb:2017iph}. This was further substantiated by LHCb with 5.9$\sigma$, a high statistical significance, observed via the  $ \Xi^{++}_{cc}\rightarrow \Xi^+_{c} \pi^+$ decay channel \cite{LHCb:2018pcs}  in 2018.
Further exploration in 2019 was made by the LHCb collaboration via a search for the $\Xi^+_{cc}$ baryon, utilizing the decay channel of  $ \Xi^{+}_{cc}\rightarrow \Lambda^+_{c} K^- \pi^+$  \cite{LHCb:2019gqy}. This particular search’s findings were later merged with those from a 2021 investigation of the  $ \Xi^{+}_{cc}\rightarrow \Xi^{+}_{c} \pi^- \pi^+ $ decay channel. The consolidated data yielded a peak significance of 4.0 standard deviations around a mass of 3620 $ \mathrm {MeV}$ for the $\Xi^{+}_{cc}$, with systematic errors incorporated \cite{LHCb:2021eaf}. This mass measurement aligns closely with the known mass of the $ \Xi_{cc}^{++} $. Despite ongoing experimental efforts to identify additional members of these baryon families, the discovery of another distinct doubly heavy baryon has not yet been achieved \cite{LHCb:2021xba}.

A significant number of researchers have dedicated efforts to investigating different attributes of these baryons, such as their masses and residues \cite{ShekariTousi:2024mso, Ebert:2002ig,Zhang:2008rt,Wang:2010hs,Lu:2017meb,Rahmani:2020pol,Yao:2018ifh,Aliyev:2022rrf,Aliev:2012iv,Aliev:2019lvd,Aliev:2012ru,Padmanath:2019ybu,Brown:2014ena,Giannuzzi:2009gh,Shah:2017liu,Shah:2016vmd,Yoshida:2015tia,Li:2022ywz,Wang:2010it,Ortiz-Pacheco:2023kjn,Wang:2018lhz,Bagan:1992za,Alrebdi:2022lat,Wang:2010vn,ShekariTousi:2026efp}, chiral effective Lagrangian \cite{Qiu:2020omj}, as well as the determination of strong coupling constants \cite{Olamaei:2021hjd,Aliev:2021hqq,Aliev:2020lly,Alrebdi:2020rev,Rostami:2020euc,Olamaei:2020bvw,Aliev:2020aon}. Further studies have focused on their strong interaction dynamics and decay mechanisms \cite{Azizi:2020zin,Qin:2021dqo,Xiao:2017dly}, radiative transitions \cite{Aliev:2021hqq,Xiao:2017udy,Lu:2017meb,Rahmani:2020pol,Li:2017pxa,Ortiz-Pacheco:2023kjn}, weak decay channels \cite{ShekariTousi:2025fjf,Tousi:2024usi,Gerasimov:2019jwp,Wang:2017mqp,Zhao:2018mrg,Xing:2018lre,Jiang:2018oak,Gutsche:2019wgu,Gutsche:2019iac,Ke:2019lcf,Cheng:2020wmk,Hu:2020mxk,Li:2020qrh,Han:2021gkl,Wang:2017azm,Shi:2017dto,Zhang:2018llc,Ivanov:2020xmw,Shi:2020qde,Hu:2017dzi,Li:2018epz,Shi:2019hbf,Shi:2019fph,Sharma:2017txj,Patel:2024mfn,Gutsche:2017hux,Gutsche:2018msz}, magnetic properties \cite{Ozdem:2018uue,Ozdem:2019zis}, lifetimes \cite{Berezhnoy:2018bde}, and mixing angles \cite{Aliev:2012nn}, applying a variety of theoretical methods. To obtain precise predictions for these parameters, nonperturbative approaches such as the QCD sum rules formulated by Shifman, Vainshtein, and Zakharov in 1979 \cite{Shifman:1978bx, Shifman:1978by} play a crucial role. This framework employs the QCD Lagrangian and relies on correlation functions constructed with different interpolating currents. The QCD sum rule method has yielded numerous reliable estimates for hadronic features, corroborated by experimental findings, and remains a powerful tool in hadron physics \cite{Aliev:2010uy,Aliev:2009jt,Aliev:2012ru,Agaev:2016dev,Azizi:2016dhy}.

Numerous theoretical frameworks have been employed to explore the spectroscopic properties of doubly heavy spin-$\frac{1}{2}$ baryons. These include lattice QCD techniques \cite{Padmanath:2019ybu,Brown:2014ena}, relativistic and nonrelativistic quark models \cite{Ebert:2002ig,Yoshida:2015tia}, the hypercentral constituent quark model \cite{Shah:2016vmd,Shah:2017liu}, QCD sum rule methods \cite{ShekariTousi:2024mso,Aliev:2012ru,Aliev:2019lvd,Zhang:2008rt,Wang:2010it,Wang:2010hs},  Bethe–Salpeter formalism \cite{Li:2022ywz}, and the Salpeter equation framework \cite{Giannuzzi:2009gh}. 

On the other hand, the study of the mass of the doubly heavy spin-$\frac {3}{2} $ baryons is done in the literature. The ground state mass determination for these baryons  have been carried out using a variety of theoretical frameworks, including lattice QCD simulations \cite{Brown:2014ena}, constituent quark models \cite{Yoshida:2015tia}, relativistic quark models \cite{Ebert:2002ig}, the QCD sum rules approach \cite{Aliev:2012iv,Bagan:1992za,Zhang:2008rt,Wang:2010vn}, hypercentral constituent quark models \cite{Shah:2016vmd,Shah:2017liu}, and solutions to the Salpeter equation \cite{Giannuzzi:2009gh}. To improve the precision and minimize uncertainties, we expand the analysis to include nonperturbative operators with dimensions reaching ten. Earlier studies using sum rules \cite{Aliev:2012iv,Bagan:1992za,Zhang:2008rt,Wang:2010vn} were limited to nonperturbative operators up to dimension five and six.

 Various theoretical methods have been employed to analyze the mass  of doubly heavy baryons with spin-$\frac{3}{2}$ in the $1P$ excited configuration, including the hypercentral constituent quark model \cite{Shah:2016vmd,Shah:2017liu}, constituent quark models \cite{Yoshida:2015tia}, and the QCD sum rules \cite{Alrebdi:2022lat,Wang:2010it}. The mass of $2S$ excited state of these baryons has also been computed in various frameworks such as the hypercentral constituent quark model \cite{Shah:2016vmd,Shah:2017liu}, constituent quark models \cite{Yoshida:2015tia}, and the QCD sum rule technique \cite{Alrebdi:2022lat}. In this work, we aim to achieve improved precision in calculating the masses of all baryon members in these excited states by incorporating higher-dimensional nonperturbative contributions, as outlined above.
We calculate the residues of doubly heavy spin-$\frac{3}{2}$ baryons across the ground state, $1P$ and $2S$ excited states, with  a  high  accuracy and up to ten  dimensional nonperturbative operator in the  operator product expansion (OPE) as well.  The ground states have been analyzed using the QCD sum rule method in Refs. \cite{Aliev:2012iv,Wang:2010vn,Bagan:1992za}. The residues corresponding to the $1P$ and $2S$  states were determined using QCD sum rules, as detailed in Refs. \cite{Alrebdi:2022lat,Wang:2010it} and Ref. \cite{Alrebdi:2022lat}, respectively.
We extend these studies by focusing on the mass spectra and residues for the spin-$\frac{3}{2}$ doubly heavy baryons in  the ground state as well as the initial radial and orbital excited configurations, which serve as essential inputs for investigating the decay mechanisms and interaction dynamics involving such baryons. The findings  may serve as valuable input for upcoming experimental investigations of such baryons at particle physics facilities like the LHC.
  The prospect of soon detecting new doubly heavy baryons motivates theoretical explorations that could significantly advance our knowledge of heavy quark dynamics within hadrons. These investigations serve as a powerful means to bridge the gap between perturbative and nonperturbative QCD, providing essential theoretical groundwork for future experimental findings. Furthermore,  doubly heavy baryons research may unlock deeper insights into the fundamental nature of other heavy baryon systems.
 
 The subsequent sections detail the methodology and findings of this study. Section II  focuses on the formulation of QCD sum rule method applied to determine the masses and residues of spin-$\frac{3}{2}$ doubly heavy baryons, specifically focusing on $	\Xi_{cc}^*$, $	\Xi_{bc}^*$, $	\Xi_{bb}^*$, $	\Omega_{cc}^*$, $	\Omega_{bc}^*$ and $	\Omega_{bb}^*$. Section III presents a comprehensive numerical analysis of these sum rules, including a discussion of the resulting mass and residue values. Furthermore, this section provides a comparative assessment of our findings in relation to predictions obtained from another theoretical studies. Finally, Section IV offers our concluding remarks and we also present some expressions in the appendix.

\section{QCD sum rules for the physical quantities of the doubly heavy baryons}\label{II}
Using the QCD sum rule method, we extract sum rules for the masses and residues of doubly heavy spin-$\frac{3}{2}$ baryons in their ground, first orbital, and first radial excitation. In table~\ref{table:1}, the features of the doubly heavy baryons  in the ground state, are listed:

\begin{table}[htb!]\caption{Doubly heavy baryons' features in the ground state.}
	\centerline{\begin{tabular}{|c|c|c|}\hline
			Baryon              & quark content        &$J^{P}$                  \\ \hline\hline
			$\Xi_{QQ}$          &$\{QQ\}q$             &$\frac{1}{2}^{+} $      \\
			\hline
			$\Xi_{QQ}^{*}$      &$\{QQ\}q$             &$\frac{3}{2}^{+} $                    \\
			\hline
			$\Omega_{QQ}$       &$\{QQ\}s$             &$\frac{1}{2}^{+} $      \\
			\hline
			$\Omega_{QQ}^{*}$   &$\{QQ\}s$             &$\frac{3}{2}^{+} $       \\
			\hline
			$\Xi_{QQ'}$         &$\{QQ'\}q$            &$\frac{1}{2}^{+} $      \\
			\hline
			$\Xi_{QQ'}^{*}$     &$\{QQ'\}q$            &$\frac{3}{2}^{+} $         \\
			\hline
			$\Omega_{QQ'}$      &$\{QQ'\}s$            &$\frac{1}{2}^{+} $    \\
			\hline
			$\Omega_{QQ'}^{*}$  &$\{QQ'\}s$            &$\frac{3}{2}^{+} $             \\
			\hline
			$\Xi_{QQ'}^{'}$     &$[QQ']q$              &$\frac{1}{2}^{+} $            \\
			\hline
			$\Omega_{QQ'}^{'}$  &$[QQ']s$              &$\frac{1}{2}^{+} $    \\
			\hline\hline
	\end{tabular}}
	\label{table:1}
\end{table}
Here, $Q$ and $Q'$ represent heavy quarks like bottom ($b$) or charm ($c$), while $q$ stands for a light quark, for example, up ($u$) or down ($d$). As a preliminary matter, we address the quark model classification of ground state doubly heavy baryons. 
For baryons like $	\Xi_{cc}^{(*)}$, $	\Xi_{bb}^{(*)}$, $	\Omega_{cc}^{(*)}$ and $	\Omega_{bb}^{(*)}$, containing two identical heavy quarks, these quarks form a diquark with total spin 1. We use the notation that starred baryons represent spin-$\frac {3}{2} $ states, with unstarred baryons having spin $\frac {1}{2} $. The coupling of the light quark (spin-$\frac {1}{2} $) with the heavy diquark (spin-1) results in total spin states of $\frac {1}{2} $ and $\frac {3}{2} $. The interpolating currents for these states must exhibit symmetry under the interchange of heavy quark fields. When considering states with two distinct heavy quarks, an additional possibility arises: beyond the formation of a diquark with total spin 1, the diquark can also possess a total spin of zero, thereby resulting in a total spin of $\frac {1}{2} $ for these states. The interpolating currents of these configurations, denoted like $	\Xi_{bc}^{'}$ and $	\Omega_{bc}^{'}$, are antisymmetric with respect to heavy quark field exchange. As previously noted, the present work focuses exclusively on spin-$\frac {3}{2} $ states.

 The analysis begins by constructing a two-point correlation function, which serves as fundamental component for obtaining hadronic parameters: 
\begin{equation}
	\Pi_{\mu\nu}(q)=i\int d^{4}xe^{iq\cdot x}\langle 0|\mathcal{T}\{\eta_{\mu}(x)\bar{\eta}_{\nu}(0)\}|0\rangle,\label{eq:CorrF1}
\end{equation}        
where $\eta_{\mu}(x)$ is an operator, known as the interpolating current, and constructed from quark field combinations. Its role is to generate the hadronic state under investigation when acting on the vacuum. The symbol $\mathcal{T}$ denotes the time ordering of field operators within a correlation function and $q$ is the doubly heavy spin-$\frac {3}{2} $ baryons four momentum.
The interpolating current of the considered doubly heavy baryons are

\begin{align}\label{eq:CorrF2}
\eta_\mu = {1\over \sqrt{3}} \epsilon^{abc} \Big\{
(q^{aT} C \gamma_\mu Q^b) Q^{\prime c} +
(q^{aT} C \gamma_\mu Q^{\prime b}) Q^c +
(Q^{aT} C \gamma_\mu Q^{\prime b}) q^c \Big\}~,
\end{align}
where  $a$, $b$ and $c$ are color indices and the charge conjugation matrix is indicated by $C$. The interpolating currents are carefully constructed to reflect the full set of quantum numbers associated with the baryonic states being investigated.

To extract physical observables within the QCD sum rule framework, we equate the theoretical (QCD) and phenomenological (physical) sides of the calculation. 
At long distances in the time-like domain of the light cone, the correlation function can be expressed in terms of physical observables such as the mass and residue of the hadronic state. This formulation is known as the physical side of the correlation function. On the other hand, the same two-point correlation function is evaluated using the OPE in the deep Euclidean region, where $q^2 \ll 0$, corresponding to the space-like sector. In this regime, the correlation function is expanded in terms of local operators with increasing mass dimensions, including contributions from quark and gluon condensates, thereby incorporating short-distance quantum effects. This formulation is referred to as the QCD side. To derive QCD sum rules for the physical quantities of interest, both representations are connected through a dispersion relation, under the assumption of quark-hadron duality. To minimize the effects of higher excited states and continuum contributions, Borel transformation and continuum subtraction techniques are employed on both sides. The final expression involves several Lorentz structures, and by equating the coefficients of identical structures, we can achieve the relevant physical parameters.

 \subsection{Physical side} 

On the physical side, the correlation function is expressed in terms of hadronic degrees of freedom. By inserting two complete sets of physical states corresponding to the quantum numbers of the initial and final baryons at the appropriate points in the calculation

\begin{eqnarray} \label{compelet set}
	1=\vert 0\rangle\langle0\vert +\sum_{h}\int\frac{d^4p_h}{(2\pi)^4}(2\pi) \delta(p^2_h-m^2_h)|h(p_h)\rangle  \langle h(p_h)|+\mbox{higher Fock states},
\end{eqnarray}
where $h(p_h)$ is the considered hadronic state with momentum $p_h$,
the hadronic contribution is then obtained through a Fourier transform and integration over the space-time coordinates, yielding an expression in momentum space

 \begin{align}
	\Pi^{Phys}_{\mu\nu}(q)&=\frac{\langle0|\eta_\mu|B_{QQ'}(q,s)\rangle\langle B_{QQ'}(q,s)|\bar{\eta_\nu}|0\rangle}{m^2-q^2}
	+\frac{\langle0|\eta_\mu|\tilde{B}_{QQ'}(q,s)\rangle\langle\tilde{B}_{QQ'}(q,s)|\bar{\eta_\nu}|0\rangle}{\tilde{m}^2-q^2}
	\nonumber\\
	&+\frac{\langle0|\eta_\mu| B_{QQ'}'(q,s)\rangle\langle B_{QQ'}'(q,s)|\bar{\eta_\nu}|0\rangle}{m'{}^2-q^2} +\cdots.
	\label{Eq:cor:Phys}
\end{align}
The states $|B_{QQ'}(q,s)\rangle$, $|\tilde{B}_{QQ'}(q,s)\rangle$, and $|B_{QQ'}'(q,s)\rangle$ correspond, respectively, to the one-particle configurations of doubly heavy baryons in the ground state with positive parity, the first orbital excitation $1P$ possessing negative parity, and the initial radial excitation $2S$ exhibiting positive parity. The notation $\cdots$ represents contributions arising from higher excited states and the continuum spectrum. We assign the symbols $m$, $\tilde{m}$, and $m'$ to denote the masses of the ground state, first orbitally excited state, and first radially excited state, respectively. To facilitate the evaluation of matrix elements appearing in Eq.~(\ref{Eq:cor:Phys}), we  introduce:
\begin{eqnarray}
	\langle 0|\eta_\mu|B_{QQ'}(q,s)\rangle&=&\lambda_{(\frac{3}{2})} u_\mu(q,s),\nonumber\\
	\langle 0|\eta_\mu|\tilde{B}_{QQ'}(q,s)\rangle&=&\tilde{\lambda}_{(\frac{3}{2})}\gamma_5 u_\mu(q,s),\nonumber\\
	\langle 0|\eta_\mu|B_{QQ'}'(q,s)\rangle&=&\lambda'_{(\frac{3}{2})} u_\mu(q,s).
\end{eqnarray}  
The parameters   $\lambda_{(\frac{3}{2})}$, $\tilde{\lambda}_{(\frac{3}{2})}$ and $\lambda'_{(\frac{3}{2})}$  represent the residues of the corresponding states of baryons. Here, 
$u_{\mu}(q,s)$ 
denotes the Rarita–Schwinger spinor corresponding to spin $s$. To advance the calculation, a summation over the Rarita-Schwinger spinors for spin-$\frac {3}{2} $ baryons is required, yielding the following expressions:
\begin{eqnarray}
	\langle 0|\eta_{\mu}|B_{QQ'}(q,s)\rangle&=&\lambda_{(\frac{1}{2})} \Bigg( {4 q_\mu \over m_{(\frac{1}{2})^+}} + \gamma_\mu \Bigg) \gamma_5 u(q,s),\nonumber\\
	\langle 0|\eta_{\mu}|\tilde{B}_{QQ'}(q,s)\rangle&=&\tilde{\lambda}_{(\frac{1}{2})} \Bigg( {- 4 q_\mu \over \tilde{m}_{(\frac{1}{2})^-}} +
	\gamma_\mu \Bigg)  u(q,s),\nonumber\\
	\langle 0|\eta_{\mu}|B_{QQ'}'(q,s)\rangle&=&\lambda'_{(\frac{1}{2})}  \Bigg( {4 q_\mu \over m'_{(\frac{1}{2})^+}} + \gamma_\mu \Bigg) \gamma_5 u(q,s).
	\label{Eq:Matrixelm2}
\end{eqnarray} 	
Here,  $\lambda_{(\frac{1}{2})}$, $\tilde{\lambda}_{(\frac{1}{2})}$, and $\lambda'_{(\frac{1}{2})}$ are introduced to represent the spin-$\frac {1}{2} $ residues corresponding to the $1S$, $1P$, and $2S$ states, respectively. The Dirac spinor is denoted by $u(q,s)$.  Analysis of the matrix elements in Eq.~\ref{Eq:Matrixelm2} reveals that Lorentz structures involving $\gamma_{\mu}$ and $q_{\mu}$ receive contributions from these spin-$\frac {1}{2} $ states. To isolate the pure spin-$\frac {3}{2} $ contributions and eliminate the contamination from spin-$\frac {1}{2} $ states, we specifically select Lorentz structures that receive contributions only from spin-$\frac {3}{2} $ states. Applying these selection criteria and the relations outlined previously, the hadronic side is then expressed as:

	\begin{eqnarray}
	\label{PhysSide1}
	\Pi_{\mu\nu}^{Phys}(q)&=&\frac{\lambda_{(\frac{3}{2})}^{2}}{m_{(\frac{3}{2})}^{2}-q^{2}}(\!\not\!{q} + m_{(\frac{3}{2})})\Big[g_{\mu\nu} -\frac{1}{3} \gamma_{\mu} \gamma_{\nu} - \frac{2q_{\mu}q_{\nu}}{3m_{(\frac{3}{2})}^{2}} +\frac{q_{\mu}\gamma_{\nu}-q_{\nu}\gamma_{\mu}}{3m_{(\frac{3}{2})}} \Big]\nonumber \\
	&+&\frac{\tilde{\lambda}_{(\frac{3}{2})}^{2}}{\tilde{m}_{(\frac{3}{2})}^{2}-q^{2}}(\!\not\!{q} -\tilde{m}_{(\frac{3}{2})})\Big[g_{\mu\nu} -\frac{1}{3} \gamma_{\mu} \gamma_{\nu} - \frac{2q_{\mu}q_{\nu}}{3\tilde{m}_{(\frac{3}{2})}^{2}} -\frac{q_{\mu}\gamma_{\nu}-q_{\nu}\gamma_{\mu}}{3\tilde{m}_{(\frac{3}{2})}} \Big]\nonumber\\
	&+&\frac{\lambda'{}^{2}}{m'{}^{2}_{(\frac{3}{2})}-q^{2}}(\!\not\!{q} + m'_{(\frac{3}{2})})\Big[g_{\mu\nu} -\frac{1}{3} \gamma_{\mu} \gamma_{\nu} - \frac{2q_{\mu}q_{\nu}}{3m'{}^{2}_{(\frac{3}{2})}} +\frac{q_{\mu}\gamma_{\nu}-q_{\nu}\gamma_{\mu}}{3m'_{(\frac{3}{2})}} \Big]+\cdots.
\end{eqnarray}	
The expression derived reveals a multitude of Lorentz structures. However, only the $g_{\mu\nu}$ and $\!\not\!{q} g_{\mu\nu}$ structures are demonstrably free from contributions arising from spin-$\frac {1}{2} $ states, originating exclusively from spin-$\frac {3}{2} $ baryon resonances. Thus, the final representation of the physical side is:
\begin{eqnarray}
	\Pi _{\mu \nu}^{phys}(q)&=&\frac{\lambda_{(\frac{3}{2})}^2}{m_{(\frac{3}{2})}^{2}-q^{2}}  (\!\not\!{q} + m_{(\frac{3}{2})})g_{\mu\nu}+
	\frac{\tilde{\lambda}_{(\frac{3}{2})}^2}{\tilde{m}_{(\frac{3}{2})}^{2}-q^{2}}  (\!\not\!{q} -\tilde{m}_{(\frac{3}{2})})g_{\mu\nu} +\frac{\lambda_{(\frac{3}{2})}^{' 2}}{m_{(\frac{3}{2})}^{'2}-q^{2}} (\!\not\!{q} + m'_{(\frac{3}{2})}) g_{\mu\nu}+\cdots.
	\label{eq:CorFun1}
\end{eqnarray}
Following the application of a Borel transformation with
respect to baryon momentum square to eliminate contributions from higher states and continuum, the result  of the correlation function of physical side is expressed in the following form:
\begin{eqnarray}
\tilde \Pi _{\mu \nu}^{phys}(q)&=&\lambda_{(\frac{3}{2})}^2 e^{-\frac{m_{(\frac{3}{2})}^{2}}{M^{2}}} (\!\not\!{q}+ m_{(\frac{3}{2})}) g_{\mu\nu}+
	\tilde{\lambda}{_{(\frac{3}{2})}}^2 e^{-\frac{\tilde{m}_{(\frac{3}{2})}^{2}}{M^{2}}}  (\!\not\!{q} -\tilde{m}_{(\frac{3}{2})})g_{\mu\nu}+\lambda_{(\frac{3}{2})}^{'2} e^{-\frac{m_{(\frac{3}{2})}^{'2}}{M^{2}}} (\!\not\!{q} + m'_{(\frac{3}{2})})g_{\mu\nu} +\cdots,
	\label{eq:CorFunBorel}
	\end{eqnarray}
where $\tilde \Pi _{\mu \nu}^{phys}(q)$ is the Borel transformed expression of the correlation function and $M^2$ is the Borel mass parameter. This transformed expression represents the physical  side of the sum rule and serves as the basis for further numerical analysis.

 \subsection{QCD side} 
From a theoretical perspective, the calculation of the two-point correlation function is performed within the deep Euclidean domain. This computation requires the substitution of the spin-$\frac{3}{2}$ doubly heavy baryon interpolating current, defined in Eq.~(\ref{eq:CorrF2}), into Eq.~(\ref{eq:CorrF1}), followed by the implementation of Wick's theorem to contract all possible quark field contractions. The outcome is an explicit expression for the correlator in terms of quark propagators:

\begin{align} \label{shekari}
\Pi_{\mu\nu} (q) &= {1\over 3} \epsilon^{abc} \epsilon^{a^\prime b^\prime
	c^\prime} \Big\{
-S_Q^{c b^\prime} \gamma_\nu \widetilde{S}_{Q^\prime}^{a a^\prime} \gamma_\mu
S_q^{b c^\prime} - S_Q^{c a^\prime} \gamma_\nu \widetilde{S}_{q}^{b b^\prime}
\gamma_\mu S_{Q^\prime}^{a c^\prime} -
S_{Q^\prime}^{c a^\prime} \gamma_\nu \widetilde{S}_{Q}^{b b^\prime} \gamma_\mu
S_{q}^{a c^\prime} \nonumber\\
& S_{Q^\prime}^{c b^\prime} \gamma_\nu \widetilde{S}_{q}^{a a^\prime}
\gamma_\mu S_{Q}^{b c^\prime} - S_q^{c a^\prime} \gamma_\nu
\widetilde{S}_{Q^\prime}^{b b^\prime} \gamma_\mu S_Q^{a c^\prime} -
S_q^{c b^\prime} \gamma_\nu
\widetilde{S}_{Q}^{a a^\prime} \gamma_\mu S_{Q^\prime}^{b c^\prime} \nonumber\\
& S_{Q^\prime}^{c c^\prime}  \mbox{Tr}\Big[S_{Q}^{b a^\prime} \gamma_\nu
\widetilde{S}_{q}^{a b^\prime} \gamma_\mu \Big] -
S_{q}^{c c^\prime}  \mbox{Tr}\Big[S_{Q^\prime}^{b a^\prime} \gamma_\nu
\widetilde{S}_{Q}^{a b^\prime} \gamma_\mu \Big] -
S_{Q}^{c c^\prime}  \mbox{Tr}\Big[S_{q}^{b a^\prime} \gamma_\nu
\widetilde{S}_{Q^\prime}^{a b^\prime} \gamma_\mu \Big] \Big\}~,
\end{align}
where $S_q$ and $S_Q$  are defined as the light and heavy quark propagators, respectively, and we also define $\widetilde{S} $  as  $CS^T C$.  The  propagators of light and heavy quarks in coordinate space  are as follows \cite{Reinders:1984sr}:

\begin{eqnarray}
	&&S_{q}^{ab}(x)=i\delta _{ab}\frac{\slashed x}{2\pi ^{2}x^{4}}-\delta _{ab}%
	\frac{m_{q}}{4\pi ^{2}x^{2}}-\delta _{ab}\frac{\langle \overline{q}q\rangle
	}{12}+i\delta _{ab}\frac{\slashed xm_{q}\langle \overline{q}q\rangle }{48}%
	-\delta _{ab}\frac{x^{2}}{192}\langle \overline{q}g_{s}\sigma Gq\rangle
	\notag \\
	&&+i\delta _{ab}\frac{x^{2}\slashed xm_{q}}{1152}\langle \overline{q}%
	g_{s}\sigma Gq\rangle -i\frac{g_{s}G_{ab}^{\alpha \beta }}{32\pi ^{2}x^{2}}%
	\left[ \slashed x{\sigma _{\alpha \beta }+\sigma _{\alpha \beta }}\slashed x%
	\right] -i\delta _{ab}\frac{x^{2}\slashed xg_{s}^{2}\langle \overline{q}%
		q\rangle ^{2}}{7776}  \notag \\
	&&-\delta _{ab}\frac{x^{4}\langle \overline{q}q\rangle \langle
		g^2_{s}G^{2}\rangle }{27648}+\cdots ,  \label{eq:A1}
\end{eqnarray}%
and
\begin{eqnarray}
	&&S_{Q}^{ab}(x)=i\int \frac{d^{4}k}{(2\pi )^{4}}e^{-ikx}\Bigg \{\frac{\delta
		_{ab}\left( {\slashed k}+m_{Q}\right) }{k^{2}-m_{Q}^{2}}-\frac{%
		g_{s}G_{ab}^{\alpha \beta }}{4}\frac{\sigma _{\alpha \beta }\left( {\slashed %
			k}+m_{Q}\right) +\left( {\slashed k}+m_{Q}\right) \sigma _{\alpha \beta }}{%
		(k^{2}-m_{Q}^{2})^{2}}  \notag  \label{eq:A2} \\
	&&+\frac{g_{s}^{2}G^{2}}{12}\delta _{ab}m_{Q}\frac{k^{2}+m_{Q}{\slashed k}}{%
		(k^{2}-m_{Q}^{2})^{4}}+\frac{g_{s}^{3}G^{3}}{48}\delta _{ab}\frac{\left( {%
			\slashed k}+m_{Q}\right) }{(k^{2}-m_{Q}^{2})^{6}}\left[ {\slashed k}\left(
	k^{2}-3m_{Q}^{2}\right) +2m_{Q}\left( 2k^{2}-m_{Q}^{2}\right) \right] \left(
	{\slashed k}+m_{Q}\right) +\cdots \Bigg \}.  \notag \\
	&&
\end{eqnarray}
Here, $G_{\mu\nu}$ is the tensor describing the gluon field strength, and $G_{ab}^{\alpha\beta}$ is defined as $G_A^{\alpha\beta} t^A_{ab}$, with the generators $t^A$ given by $\lambda^A / 2$. The quantity $G^{2}$ is expressed as the  $G_{\alpha \beta}^{A} G_{A}^{\alpha \beta}$, while $G^{3}$ involves the color structure constants $f^{ABC}$ and is written as $f^{ABC} G_{\alpha \beta}^{A} G^{B \beta \delta} G_{\delta}^{C \alpha}$. Here, $\lambda^{A}$ denote the Gell-Mann matrices, and $f^{ABC}$ are the structure constants associated with the $SU_c(3)$ color gauge group. The indices $A$, $B$, and $C$ vary from 1 to 8.

The correlation function on the QCD side is represented as a combination of perturbative and nonperturbative operators, each corresponding to distinct mass dimensions. This representation is achieved by substituting the light and heavy quark propagators, given in Eq.~(\ref{eq:A1}) and Eq.~(\ref{eq:A2}), into Eq.~(\ref{shekari}). In the evaluation, we incorporate nonperturbative effects up to dimension ten. As previously discussed, two independent Lorentz structures, namely $g_{\mu\nu}$ and $\not\!q g_{\mu\nu}$, appear on both the QCD and hadronic representations. After applying Borel transformation and performing continuum subtraction, the final expressions for the QCD side are obtained for these two Lorentz structures:
\begin{eqnarray}
	\tilde\Pi^{\mathrm{QCD}}_{\not\!q g_{\mu\nu}}(s_0,M^2)=\int_{(m_Q+m_{Q'})^2}^{s_0}ds\,e^{-\frac{s}{M^2}}\rho_{\not\!q g_{\mu\nu}}(s)+\Gamma_{\not\!q g_{\mu\nu}}(M^2),
	\label{Eq:finalCor:QCD1}
\end{eqnarray}
and
\begin{eqnarray}
	\tilde\Pi^{\mathrm{QCD}}_{ g_{\mu\nu}}(s_0,M^2)=\int_{(m_Q+m_{Q'})^2}^{s_0}ds\,e^{-\frac{s}{M^2}}\rho_{ g_{\mu\nu}}(s)+\Gamma_{ g_{\mu\nu}}(M^2),
	\label{Eq:finalCor:QCD}
\end{eqnarray}
where $s_0$ is the continuum threshold and $\rho (s)=\frac{1}{\pi}\mathrm{Im}[\tilde\Pi^{\mathrm{QCD}}(s_0,M^2)]$  denotes the spectral density for the $\not\!q g_{\mu\nu}$ and $g_{\mu\nu}$   Lorentz structures, encompassing perturbative operator and nonperturbative operators of dimension three to five. The  $\Gamma_{\not\!q g_{\mu\nu}}(M^2)$ and $\Gamma_{ g_{\mu\nu}}(M^2)$ signify a real valued functions incorporating nonperturbative operators of dimension six to ten. The comprehensive mathematical expressions for $\rho _{~\not\!q g_{\mu\nu}}(s)$ and $\rho _{ g_{\mu\nu}}(s)$  are detailed in the Appendix.  Due to its extensive form, the explicit expression for $\Gamma_{\not\!q g_{\mu\nu}}(M^2)$  is not shown in this work.
A comparative analysis of the results derived from both sides is conducted via dispersion relations, considering coefficients of identical Lorentz structures, $\not\!q g_{\mu\nu}$ and $g_{\mu\nu}$. This procedure leads to QCD sum rules governing the quantities, the mass and residue, given by: 

\begin{eqnarray}
	\lambda_{(\frac{3}{2})}^2 e^{-\frac{m_{(\frac{3}{2})}^2}{M^2}}+\tilde\lambda_{(\frac{3}{2})}^2e^{-\frac{\tilde m_{(\frac{3}{2})}^2}{M^2}}+\lambda_{(\frac{3}{2})}^{'2}e^{-\frac{m_{(\frac{3}{2})}^{'2}}{M^2}}=\tilde\Pi^{QCD}_{~\!\not\!{q} ~g_{\mu\nu}}(s_0,M^2),
	\label{Eq:cor1}
\end{eqnarray}
and
\begin{eqnarray}
	\lambda_{(\frac{3}{2})}^2 m_{(\frac{3}{2})} e^{-\frac{m_{(\frac{3}{2})}^2}{M^2}}-\tilde{\lambda}_{(\frac{3}{2})}^2\tilde{m}_{(\frac{3}{2})} e^{-\frac{\tilde{m}_{(\frac{3}{2})}^2}{M^2}}+\lambda_{(\frac{3}{2})}^{'2} m'_{(\frac{3}{2})}e^{-\frac{m_{(\frac{3}{2})}^{'2}}{M^2}}=\tilde\Pi^{QCD}_{g_{\mu\nu}}(s_0,M^2).
	\label{Eq:cor2}
\end{eqnarray}
Subsequently, we elucidate the extraction of QCD sum rules for the masses and residues of the ground  and excited states associated with the structure $ \not\!q g_{\mu\nu}$.  This objective is achieved through a detailed three-stage methodology.  Initially, the mass and residue of the ground state are independently evaluated. Consequently, we adopt the ground state + continuum approach, wherein terms beyond the leading term on the left-hand side of Eq.~(\ref{Eq:cor1}) are treated as continuum contributions. With modulating the continuum threshold, $ s_0 $, we ensure that the calculation is exclusively influenced by the ground state.  To ascertain the mass and residue of the ground state, a system of two equations is required, derived from the derivative of Eq.~(\ref{Eq:cor1}) with respect to $ -\frac{1}{M^2} $. Following algebraic manipulation and focusing solely on the leading term of the left-hand side of Eq.~(\ref{Eq:cor1}), the expressions for the mass and residue of the ground state are derived through the sum rules, which can be written as:

\begin{eqnarray}
	m_{(\frac{3}{2})}^2=\frac{\frac{d}{d(-\frac{1}{M^2})}\tilde\Pi^{\mathrm{QCD}}_{~\not\! q~ g_{\mu\nu}}(s_0,M^2)}{\Pi^{\mathrm{QCD}}_{~\not\!q ~g_{\mu\nu}}(s_0,M^2)},
	\label{Eq:mass:Gr}
\end{eqnarray}
and
\begin{eqnarray}
	\lambda_{(\frac{3}{2})}^2=e^{\frac{m^2}{M^2}}\tilde\Pi^{\mathrm{QCD}}_{~\not\!q ~g_{\mu\nu}}(s_0,M^2).
	\label{Eq:residumass:Gr}
\end{eqnarray}
Subsequently, once the mass and residue of the ground state have been determined with sufficient accuracy, the analysis proceeds by gradually increasing the continuum threshold $ s_0 $ so as to accommodate the contribution of the first orbital excitation, $1P$.
 In this instance, the initial two terms on the left-hand side of Eq.~(\ref{Eq:cor1}), corresponding to the $1S$ and $1P$ configurations, are taken into account. 
  Consequently, the third term, representing the radially excited, $2S$ state, is incorporated within the continuum, thereby implementing the ground state + first orbitally excited state + continuum scheme.  
  Within this approach, the mass and residue of the ground state serve as known inputs, while two additional unknowns appear specifically, the mass and residue of the $1P$ state. These new parameters can then be extracted in close analogy to the procedure already employed for the ground state.  Ultimately, the ground state + first orbitally excited state + first radially excited state + continuum scheme is employed to deduce the mass and residue of the radially excited $2S$ state, achieved by further incrementing the continuum threshold  $ s_0 $ and utilizing the spectroscopic parameters of the ground and $1P$ states as inputs.  These stepwise methodologies progressively extending the spectrum from the ground state to the first orbital excitation and subsequently to the first radial excitation will be systematically revisited in the next section, where detailed numerical analyses are provided to substantiate the outlined procedure.
 
\section{Numerical results and discussions }

Now, we detail the numerical evaluation of our findings concerning the masses and residues of doubly heavy spin-${(\frac{3}{2})}$ baryons, encompassing their ground, first orbitally, and first radially excited states. 

In this work, the quark masses are adopted within the $\overline{MS}$ scheme using the values reported by the Particle Data Group (PDG)~\cite{Zyla:2020zbs}: 
$m_{c}(m_c)=(1.27\pm0.02)\,\mathrm{GeV}$, 
$m_{b}(m_b)=(4.18\pm0.03)\,\mathrm{GeV}$, and 
$m_s(\mu=2\,\mathrm{GeV})=(0.095\pm0.005)\,\mathrm{GeV}$. 
The variation of these masses with respect to the energy scale is determined through the renormalization group equations.
In our analysis, the vacuum condensates are taken at the typical reference scale $\mu = 1\,\mathrm{GeV}$ with the standard values  
$\langle \bar{q}q \rangle = - (0.24 \pm 0.01\,\mathrm{GeV})^3$,  $\langle \bar{s}s \rangle =0.8  \langle \bar{q}q \rangle $,
$\langle \bar{q} g_s \sigma G q \rangle = m_0^2 \langle \bar{q}q \rangle$,  
$m_0^2 = (0.8 \pm 0.1)\,\mathrm{GeV}^2$,   
$\langle \frac{\alpha_s GG}{\pi} \rangle = (0.012 \pm 0.004)\,\mathrm{GeV}^4$ and $\langle g_s^3 G^3 \rangle = (0.57\pm0.29)$ $~\mathrm{GeV}^6 $,  
as commonly adopted in QCD sum rule calculations~\cite{Belyaev:1982sa,Belyaev:1982cd,Narison:2015nxh}.
To account for the scale dependence of the relevant parameters, we apply the renormalization group evolution laws (see for instance~\cite{Wang:2021dcb,Wang:2014oca} and references therein):  
\begin{eqnarray}
	m_s(\mu)&=&m_s(2\,\mathrm{GeV})\left[\frac{\alpha_{s}(\mu)}{\alpha_{s}(2\,\mathrm{GeV})}\right]^{\frac{4}{9}}\,,\nonumber\\
	m_c(\mu)&=&m_c(m_c)\left[\frac{\alpha_{s}(\mu)}{\alpha_{s}(m_c)}\right]^{\frac{12}{25}}\,,\nonumber\\
	m_b(\mu)&=&m_b(m_b)\left[\frac{\alpha_{s}(\mu)}{\alpha_{s}(m_b)}\right]^{\frac{12}{23}}\,,\nonumber\\
	\langle \bar{q}q \rangle(\mu) &=& \langle \bar{q}q \rangle(1\,\mathrm{GeV})
	\left[\frac{\alpha_s(1\,\mathrm{GeV})}{\alpha_s(\mu)}\right]^{\frac{12}{33 - 2n_f}}\,, \nonumber\\
	\langle \bar{q} g_s \sigma G q \rangle(\mu) &=& \langle \bar{q} g_s \sigma G q \rangle(1\,\mathrm{GeV})
	\left[\frac{\alpha_s(1\,\mathrm{GeV})}{\alpha_s(\mu)}\right]^{\frac{2}{33 - 2n_f}}\,, \nonumber\\
	\alpha_s(\mu) &=& \frac{1}{b_0 t}
	\left[ 1 - \frac{b_1}{b_0^2}\frac{\ln t}{t}
	+ \frac{b_1^2(\ln^2 t - \ln t - 1) + b_0 b_2}{b_0^4 t^2} \right]\,, 
\end{eqnarray}
where $t = \ln\left(\frac{\mu^2}{\Lambda^2}\right)$, 
and the coefficients take the values  
$b_0 = \frac{33 - 2n_f}{12\pi}$,  
$b_1 = \frac{153 - 19n_f}{24\pi^2}$, and  
$b_2 = \frac{2857 - \frac{5033}{9}n_f + \frac{325}{27}n_f^2}{128\pi^3}$.  
The QCD scale parameter $\Lambda$ is taken to be $213\,\mathrm{MeV}$, $296\,\mathrm{MeV}$, and $339\,\mathrm{MeV}$ for the flavor numbers $n_f=5$, $4$, and $3$, respectively, following PDG conventions ~\cite{Zyla:2020zbs}. In these calculations, the masses of the $u$ and $d$ quarks are neglected.
All input parameters are therefore evolved to an appropriate energy scale $\mu$ consistent with this choice in order to determine the corresponding hadron masses and residues.
Here,  we take $n_f=4$ for the $	\Xi_{cc}^{*}$ and $	\Omega_{cc}^{*}$ baryons and  $n_f=5$ for the  $	\Xi_{bc}^{*}$,  $	\Xi_{bb}^{*}$,  $\Omega_{bc}^{*}$ and $\Omega_{bb}^{*}$ baryons.  As we have the bottom and charm quarks  in the considered states, we take the interval $ \mu$ =(2-4) GeV for these calculations. Finally, we report the averaged values of the mass and the residue obtained for $\mu$= 2, 3 and 4 GeV. The gluon condensate values  
$\langle \frac{\alpha_s GG}{\pi} \rangle = (0.012 \pm 0.004)\,\mathrm{GeV}^4$  and $\langle g_s^3 G^3 \rangle = (0.57 \pm0.29)$ $~\mathrm{GeV}^6 $
remain widely used in contemporary QCD sum rule studies, as it continues to represent the most reliable estimate in the absence of more precise determinations.

In the present work, the operator product expansion (OPE) is carried out consistently up to vacuum condensates of dimension 10. The higher-dimensional condensates are expressed in terms of lower-dimensional ones using the factorization assumption.  
As discussed by Shifman, Vainshtein, Zakharov, and Ioffe~\cite{Shifman:1978bx, Shifman:1978by,Ioffe:2005ym}, 
the factorization (or vacuum saturation) hypothesis is expected to hold with an accuracy of order $1/N_c^2$. 
In the theoretical limit of large $N_c$, the correction term $1/N_c^2$ tends to zero, indicating that the approximation becomes exact. 
Nevertheless, for the physical case of $N_c = 3$, the suppression factor is finite, 
$1/N_c^2 \approx 10\%$, which defines the typical uncertainty associated with factorization.
Consequently, violations of factorization represent an intrinsic source of theoretical uncertainty in QCD sum rule calculations involving multi-quark condensates. To parametrize possible non-factorizable contributions in a systematic and transparent way, we introduce a phenomenological parameter $\kappa$ that measures deviations from vacuum saturation ~\cite{Wang:2021dcb}.  In particular, the factorized condensate structures are modified as
\begin{align}
	\langle \bar q q \rangle^2 
	&\rightarrow 
	\kappa\, \langle \bar q q \rangle^2 , \nonumber\\
	\langle \bar q q \rangle 	\left\langle \frac{\alpha_s}{\pi} G G \right\rangle
	&\rightarrow 
	\kappa\, 	\langle \bar q q \rangle 	\left\langle \frac{\alpha_s}{\pi} G G \right\rangle , \nonumber\\
	\langle \bar q q \rangle \langle \bar q g_s \sigma G q \rangle 
	&\rightarrow 
	\kappa\, \langle \bar q q \rangle \langle \bar q g_s \sigma G q \rangle , \nonumber\\
	\left\langle \frac{\alpha_s}{\pi} G G \right\rangle  \langle \bar q g_s \sigma G q \rangle
	&\rightarrow 
	\kappa\, 	\left\langle \frac{\alpha_s}{\pi} G G \right\rangle  \langle \bar q g_s \sigma G q \rangle , \nonumber\\
	\langle \bar q g_s \sigma G q \rangle^2 
	&\rightarrow 
	\kappa\, \langle \bar q g_s \sigma G q \rangle^2 ,\nonumber\\
	\langle \bar q q \rangle^2 
	\left\langle \frac{\alpha_s}{\pi} G G \right\rangle
	&\rightarrow 
	\kappa\, \langle \bar q q \rangle^2 
	\left\langle \frac{\alpha_s}{\pi} G G \right\rangle ,\cdots .
\end{align}
Here, $\kappa=1$ corresponds to exact vacuum saturation, while $\kappa>1$ accounts for possible non-factorizable effects. Several determinations of the parameter $\kappa$ have been reported in the literature, with its magnitude varying according to the specific channel under investigation. 
The available estimates span a broad interval, starting from $\kappa = 1$~\cite{Ioffe:2010zz}, increasing to approximately $2$--$3$~\cite{Narison:2002woh}, and reaching values as large as $6$ in certain analyses~\cite{Leinweber:1995fn}.
 Although, in principle, different condensate structures may be associated with different $\kappa$ values, introducing independent parameters would significantly enlarge the parameter space without clear phenomenological guidance. For this reason, and to maintain a controlled uncertainty analysis, we adopt a common $\kappa$ for all factorized higher-dimensional condensates. To obtain a conservative and realistic estimate of the theoretical uncertainty induced by the factorization assumption, we vary the parameter within the range
\[
1 \le \kappa \le 5 .
\]
For each value of $\kappa$, the sum rules are recalculated, and the corresponding masses and residues are extracted. The maximal deviation of these quantities relative to their central values is then incorporated into the total theoretical uncertainty. The calculations resulted in an estimated uncertainty of roughly $2\%$  in the determined mass and residue of the hadrons.

\begin{figure}[h!]
	\begin{center}
		\includegraphics[totalheight=6cm,width=8cm]{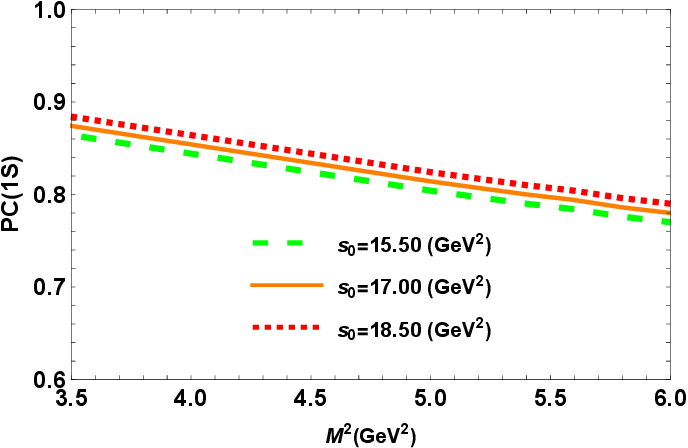}
	\end{center}
	\caption{ PC for the $	\Xi_{cc}^*$ as a function of the Borel mass parameter $M^2$ 
		at various values of the parameter $s_0$.}
	\label{fig:pcchicc}
\end{figure}

 Apart from these parameters, the QCD sum rules require two additional parameters that need to be fixed: the Borel parameter, $M^2$, and the threshold parameter, $s_0$. These results are obtained by applying the standard conditions of the QCD sum rule approach, which include minimal dependence on auxiliary parameters, dominance of the pole contribution, and the convergence of the OPE.
 OPE convergence is attained when the perturbative contribution predominates over the nonperturbative components, with higher-dimensional nonperturbative operators exhibiting progressively diminishing contributions. Pole dominance ensures that the resonances under investigation obtain a primary contribution relative to higher excited states and the continuum. Consequently, the upper bound of the $M^2$ parameter is dictated by the stipulation that the pole contributions (PC) exceed those originating from higher excited states and the continuum. According to these considerations, the subsequent conditions must be satisfied:
\begin{eqnarray}
	\mathrm{PC}=\frac{\Pi(s_0,M^2)}{\Pi(\infty,M^2)}\geq 0.5.
\end{eqnarray}
Figure~\ref{fig:pcchicc} demonstrates how the pole contribution of $\mathrm{\Xi_{cc}^*}$ satisfies this requirement, shown as a function of the Borel parameter $M^2$ for different choices of the threshold parameter $s_0$.

The lower bound for $M^2$ is established by enforcing the convergence of the OPE series. Specifically, this necessitates the dominance of the perturbative contribution over the nonperturbative terms, with nonperturbative operators of increasing dimensionality exhibiting decreasing contributions. To formalize our requirement, we consider the subsequent ratio:
\begin{equation}
	\frac{\Pi ^{\mathrm{DimN}}(s_0,M^2)}{\Pi (s_0,M^2)}\le\ 0.05.
	\label{eq:Convergence}
\end{equation} 
Here, $\mathrm{DimN}$ denotes the cumulative contribution from the highest three mass dimensions considered, specifically $\mathrm{DimN=Dim(8+9+10)}$.

\begin{figure}[h!]
	\begin{center}
		\includegraphics[totalheight=6cm,width=8cm]{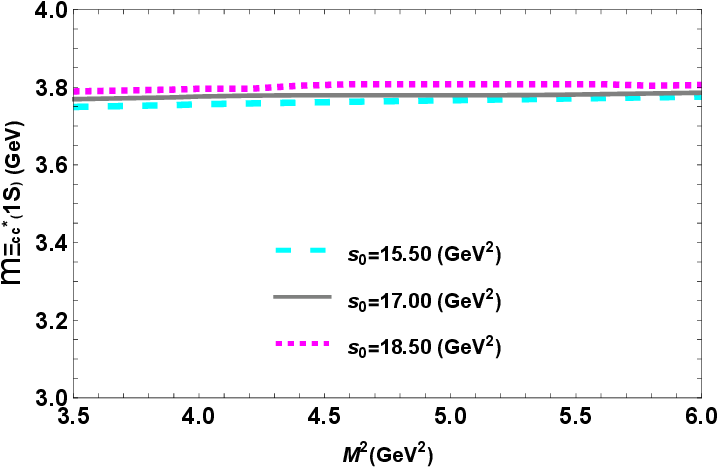}
		\includegraphics[totalheight=6cm,width=8cm]{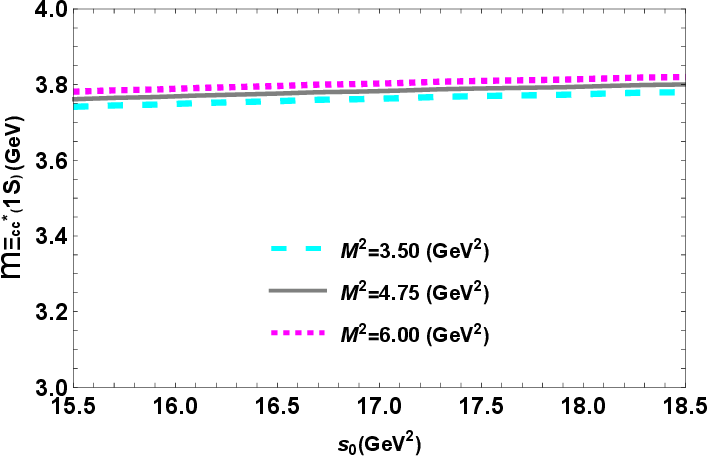}
	\end{center}
	\caption{\textbf{Left:} The $1S$ mass of $\mathrm{\Xi_{cc}^*}$ as a function of the Borel parameter $M^2$, shown for several fixed $s_0$ values.
			\textbf{Right:} The $1S$ mass of $\mathrm{\Xi_{cc}^*}$ as a function of the threshold parameter $s_0$, shown for several fixed $M^2$ values.}
	\label{gr:g1}
\end{figure}

\begin{figure}[h!]
	\begin{center}
		\includegraphics[totalheight=6cm,width=8cm]{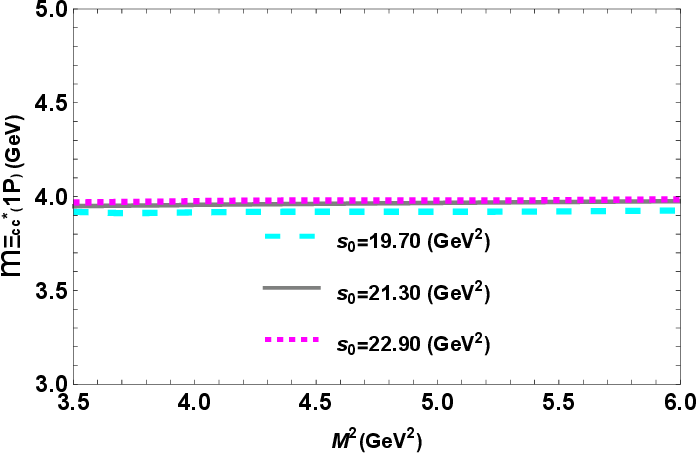}
		\includegraphics[totalheight=6cm,width=8cm]{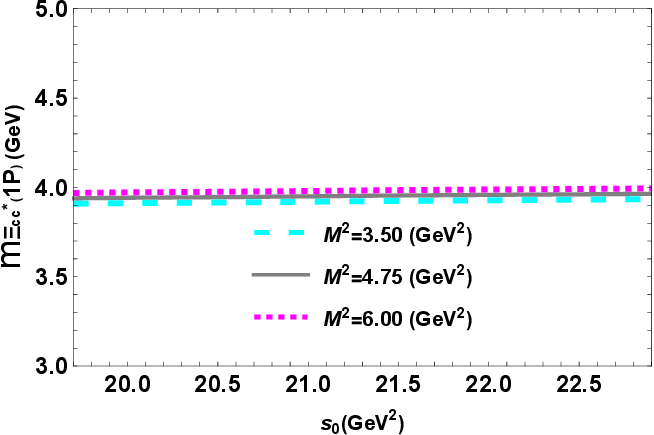}
	\end{center}
\caption{\textbf{Left:} The $1P$ mass of $\mathrm{\Xi_{cc}^*}$ as a function of the Borel parameter $M^2$, shown for several fixed $s_0$ values.
	\textbf{Right:} The $1P$ mass of $\mathrm{\Xi_{cc}^*}$ as a function of the threshold parameter $s_0$, shown for several fixed $M^2$ values.}
	\label{gr:g2}
\end{figure}

\begin{figure}[h!]
	\begin{center}
		\includegraphics[totalheight=6cm,width=8cm]{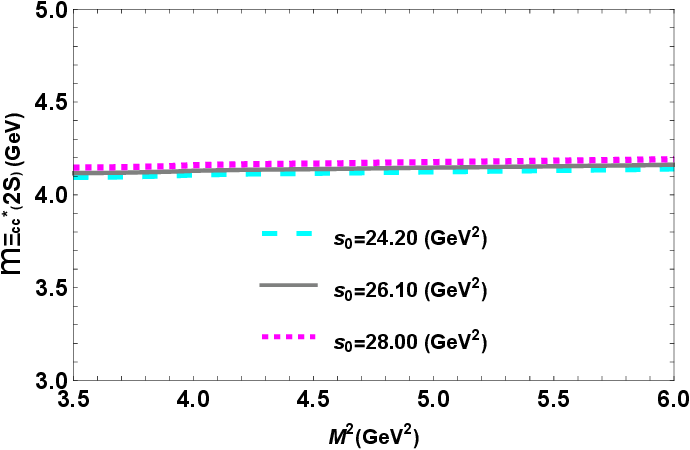}
		\includegraphics[totalheight=6cm,width=8cm]{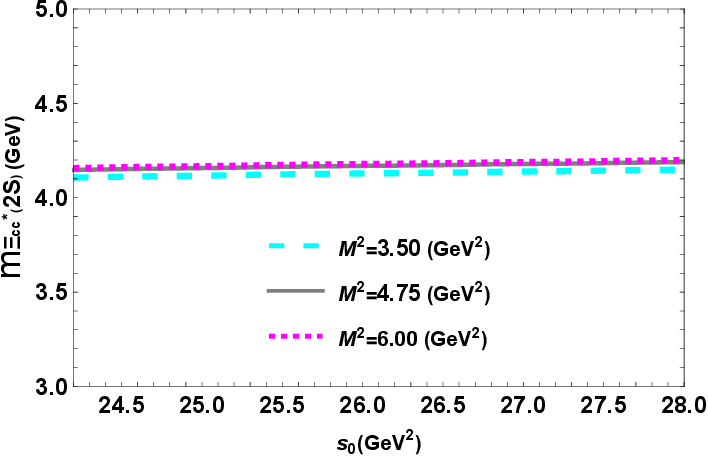}
	\end{center}
\caption{\textbf{Left:} The $2S$ mass of $\mathrm{\Xi_{cc}^*}$ as a function of the Borel parameter $M^2$, shown for several fixed $s_0$ values.
	\textbf{Right:} The $2S$ mass of $\mathrm{\Xi_{cc}^*}$ as a function of the threshold parameter $s_0$, shown for several fixed $M^2$ values.}
	\label{gr:g3}
\end{figure}

The continuum threshold parameter $s_0$, which reflects the onset of excited state contributions, is determined to ensure that the ground state dominates within the chosen analysis window. This selection prevents contamination from higher resonances and continuum states in the sum rule evaluations. The chosen intervals for both the Borel mass parameter $M^2$ and the threshold value $s_0$ for each channel are listed in table~\ref{tab:results}. The numerical analysis reveals that the physical quantities remain largely stable under changes in the auxiliary parameters within the chosen intervals for the borel mass parameter $M^2$ and continuum threshold $s_0$.
 To illustrate the dependence of the masses on these parameters, we provide Figs.~\ref{gr:g1}, \ref{gr:g2}, and \ref{gr:g3} for $  \kappa=1 $ and at average value of the scale $ \mu $. The findings indicate robust stability concerning the Borel parameter and continuum threshold across their operational ranges. Among the analyzed states, the $\Xi_{cc}^*$ baryon shows representative behavior, and thus, its graphical results are included as a demonstration. The main sources of uncertainty in the computed values arise from the uncertainties in auxiliary and input parameters.


 \begin{table}[]
 	\begin{tabular}{|c|c|c|c|c|c|}
 		\hline
 		Baryon  & State &$M^2~(\mathrm{GeV^2})$&$s_0~(\mathrm{GeV^2})$  & $\mathrm {m}~(\mathrm{GeV})$ & $\mathrm {\lambda}~(\mathrm{GeV^3})$ \\ \hline\hline
 		&$	\Xi_{cc}^*(\frac{3}{2}^+)(1S)$ &$3.50-6.00$& $3.93^2-4.30^2$& $3.68  _{-0.14} ^{+0.15}$ & $0.15 \pm0.02$  \\ \cline{2-6} 
 		$	\Xi_{cc}^*$   &$\Xi_{cc}^*(\frac{3}{2}^-)(1P)$ &$3.50-6.00$& $4.43^2-4.78^2$& $3.85 _{-0.15} ^{+0.16}$ & $0.17 \pm0.02$  \\ \cline{2-6} 
 		&$\Xi_{cc}^*(\frac{3}{2}^+)(2S)$ &$3.50-6.00$& $4.91^2-5.29^2$& $3.99  \pm0.16$ & $0.19  \pm0.03$ \\ \hline\hline
 		&$\Xi_{bc}^*(\frac{3}{2}^+)(1S)$ &$6.00-9.00$& $6.96^2-7.25^2$& $6.95  \pm0.20$ & $0.30  \pm0.03$ \\ \cline{2-6} 
 		$\Xi_{bc}^*$     &$\Xi_{bc}^*(\frac{3}{2}^-)(1P)$ &$6.00-9.00$& $7.46^2-7.75^2$& $7.16  \pm0.21$ & $0.34\pm0.03$ \\ \cline{2-6} 
 		&$\Xi_{bc}^*(\frac{3}{2}^+)(2S)$ &$6.00-9.00$& $7.96^2-8.25^2$& $7.37  \pm0.21$ & $0.36_{-0.03} ^{+0.04}$ \\ \hline\hline
 		&$	\Xi_{bb}^*(\frac{3}{2}^+)(1S)$ &$10.00-15.00$& $10.68^2-10.96^2$& $10.28  \pm0.30$ & $0.50 \pm0.05$  \\ \cline{2-6} 
 		$	\Xi_{bb}^*$   &$\Xi_{bb}^*(\frac{3}{2}^-)(1P)$ &$10.00-15.00$& $11.18^2-11.46^2$& $10.48  \pm0.32$ & $0.67 _{-0.05} ^{+0.06}$  \\ \cline{2-6} 
 		&$\Xi_{bb}^*(\frac{3}{2}^+)(2S)$ &$10.00-15.00$& $11.68^2-11.96^2$& $10.71 \pm0.32 $ & $0.77\pm0.07$ \\ \hline\hline
 		&$\Omega_{cc}^*(\frac{3}{2}^+)(1S)$ &$3.50-6.00$& $3.94^2-4.31^2$& $3.74  \pm0.15$ & $0.16  \pm0.02$ \\ \cline{2-6} 
 		$\Omega_{cc}^*$     &$\Omega_{cc}^*(\frac{3}{2}^-)(1P)$ &$3.50-6.00$& $4.44^2-4.79^2$& $3.96_{-0.16} ^{+0.15}$ & $0.20_{-0.02} ^{+0.03}$ \\ \cline{2-6} 
 		&$\Omega_{cc}^*(\frac{3}{2}^+)(2S)$ &$3.50-6.00$& $4.92^2-5.30^2$& $4.10_{-0.17} ^{+0.16}$ & $0.21 \pm0.03$ \\ \hline\hline
 		&$\Omega_{bc}^*(\frac{3}{2}^+)(1S)$ &$6.00-9.00$&$6.96^2-7.25^2$& $7.01_{-0.19} ^{+0.20}$ & $0.33  \pm0.04$ \\ \cline{2-6} 
 		$\Omega_{bc}^*$     &$\Omega_{bc}^*(\frac{3}{2}^-)(1P)$ &$6.00-9.00$& $7.46^2-7.75^2$& $7.21  \pm0.21$ & $0.36 _{-0.04} ^{+0.05}$ \\ \cline{2-6} 
 		&$\Omega_{bc}^*(\frac{3}{2}^+)(2S)$ &$6.00-9.00$& $7.96^2-8.25^2$& $7.40  \pm0.21$ & $0.38  \pm0.05$ \\ \hline\hline
 		&$\Omega_{bb}^*(\frac{3}{2}^+)(1S)$ &$10.00-15.00$& $10.68^2-10.96^2$& $10.34\pm0.30$ & $0.53\pm0.05$ \\ \cline{2-6} 
 		$\Omega_{bb}^*$     &$\Omega_{bb}^*(\frac{3}{2}^-)(1P)$ &$10.00-15.00$& $11.18^2-11.46^2$& $10.60 \pm0.32$ & $0.70  \pm0.06$ \\ \cline{2-6} 
 		&$\Omega_{bb}^*(\frac{3}{2}^+)(2S)$ &$10.00-15.00$& $11.68^2-11.96^2$& $10.82  \pm0.32$ & $0.80 \pm0.08$ \\ \hline\hline
 	\end{tabular}
 	\caption{Auxiliary parameters with the computed masses and residues of the ground state, as well as the first orbital and radial excitations.
 	}
 	\label{tab:results}
 \end{table}
 
After identifying appropriate working regions of both the Borel mass parameter $M^2$ and the threshold value $s_0$, we extract the masses and residues of these considered baryons  for the ground and excited states. Starting from Eq.~(\ref{Eq:cor1}), the masses and residues of the ground state doubly heavy baryons are determined step by step, employing an optimized value of the threshold parameter under the ground state plus continuum approach.
The suitable intervals for the continuum threshold $s_0$ is identified and presented in table~\ref{tab:results}. Next, this procedure is extended to include both the ground and first orbitally excited states alongside the continuum, with appropriately chosen threshold parameters, and the resulting values are also summarized in table~\ref{tab:results}. Finally, the analysis incorporates the first radially excited $2S$ state by applying a scheme encompassing  the $1S$, $1P$, and $2S$ states and continuum, with threshold parameters adjusted accordingly. The corresponding results for the $2S$ states are included in table~\ref{tab:results}. It is noteworthy that calculations based on two different Lorentz structures $\!\not\!{q} g_{\mu\nu}$ and $ g_{\mu\nu}$ yield comparable outcomes; hence, we present only the findings derived from the $\!\not\!{q} g_{\mu\nu}$ structure.

As presented in table~\ref{tab:results}, the masses of the first orbital $1P$ and radial $2S$ excited states of $\Xi_{cc}^*$ exceed that of the ground state by approximately 0.17 and 0.31~$\mathrm{GeV}$, respectively. Corresponding mass differences for the other baryons are: $\Xi_{bc}^*$ (0.21 and 0.42~$\mathrm{GeV}$), $\Xi_{bb}^*$ (0.20 and 0.43~$\mathrm{GeV}$), $\Omega_{cc}^*$ (0.22 and 0.36~$\mathrm{GeV}$), $\Omega_{bc}^*$ (0.20 and 0.39~$\mathrm{GeV}$), and $\Omega_{bb}^*$ (0.26 and 0.48~$\mathrm{GeV}$).

Table~\ref{tab1} provides a comprehensive comparison of the predicted masses for the spin-$\frac{3}{2}$ doubly heavy  baryons in the $1S$ state, derived using various theoretical frameworks. Our analysis incorporates nonperturbative operators with mass dimensions extended up to 10, aiming to enhance precision and minimize uncertainties.  As can be seen in table~\ref{tab1}, our predictions are in good agreement with the Lattice QCD results \cite{Brown:2014ena}. The mass estimates obtained here show consistency with results from alternative techniques, including the relativistic quark model \cite{Ebert:2002ig}, QCD sum rule method \cite{Aliev:2012iv}, and constituent quark models \cite{Yoshida:2015tia}, within the stated error margins for several channels.

\begin{table}[tbp]
	\centering
	\begin{tabular}{|c|c|c|c|c|c|c|c|}
		
		\hline   Baryon& Our work & \cite{Aliev:2012iv} & \cite{Brown:2014ena}&  \cite{Ebert:2002ig} &   \cite{Zhang:2008rt} & \cite{Yoshida:2015tia} &    \cite{Wang:2010vn}  \\ \hline\hline
		$	\Xi_{cc}^*$& $3.68  _{-0.14} ^{+0.15}$                         &$3.69  \pm0.16$&$3.692\pm0.028\pm0.021$&$3.727$&	$3.90 \pm 0.10$&$3.754$&$3.61\pm0.18$	  \\  
		$	\Xi_{bc}^*$&$6.95 \pm0.20$                                   &$7.25  \pm0.20$&$6.985\pm0.036\pm0.028$&$6.98$&	$8.00 \pm 0.26$&$-$&$-$	 \\ 
		$	\Xi_{bb}^*$&$10.28  \pm0.30$                 &$10.4  \pm 1.0$&$10.187\pm0.030\pm0.024$&	$10.237 $&$10.35  \pm 0.08  $ &$10.339	$&$10.22\pm0.15$  \\
		$	\Omega_{cc}^*$&$3.74  \pm0.15$                              & $3.78  \pm0.16$&$3.822\pm0.030\pm0.024$&$3.872$ &	$3.81 \pm 0.06 $&$3.883$&$3.76 \pm0.17$	 \\ 
		$	\Omega_{bc}^*$&$7.01_{-0.19} ^{+0.20}$   & $7.3  \pm0.2$&$7.059\pm0.028\pm0.021$&$7.13 $&$7.54\pm0.08 $ &$-$	&$-$	\\
		$	\Omega_{bb}^*$	&$10.34\pm0.30$              &$10.5  \pm0.2$&$10.308\pm0.027\pm0.021$&$10.389$&$10.28 \pm0 0.05  $ & $10.467	$	& $10.38\pm0.14$\\ 
		\hline\hline
	\end{tabular}%
	\caption{Ground state $1S$ masses (in $\mathrm{GeV}$) of doubly heavy baryons and comparison with other existing theoretical predictions.}
	\label{tab1}
\end{table}   

As previously mentioned, the mass spectra of the spin-$\frac{3}{2}$  doubly heavy baryons in the $1P$ and $2S$  states have been explored using various theoretical approaches. Our study provides mass predictions for these excited states with improved precision. The results displayed in tables~\ref{tab3} and~\ref{tab5} demonstrate that, for most channels, our findings align well with those obtained from different frameworks, including the hypercentral constituent quark model \cite{Shah:2016vmd,Shah:2017liu}, constituent quark models \cite{Yoshida:2015tia}, and the QCD sum rule method \cite{Wang:2010it}, all within the quoted uncertainties.

\begin{table}[tbp]
	\centering
	\begin{tabular}{|c|c|c|c|c|c|c|}
		
		\hline Baryon& Our work &  \cite{Shah:2017liu} &  \cite{Shah:2016vmd} &   \cite{Wang:2010it}  &  \cite{Yoshida:2015tia} &   \cite{Alrebdi:2022lat} \\ \hline\hline
		$	\Xi_{cc}^*$& $3.85 _{-0.15} ^{+0.16}$     &$3.85$       &$-$&	$3.80 \pm 0.18$&$3.949$	&$5.15  \pm0.05$  \\  
		$	\Xi_{bc}^*$&$7.16  \pm0.21$                       &$7.15$            &$-$&$-$&	$-$&$7.95  \pm0.05$	 \\ 
		$	\Xi_{bb}^*$&$10.48  \pm0.32$  &$10.51$&$-$&$10.43 \pm  0.15  $&$10.476	$	&$10.95  \pm0.025$ \\
		$	\Omega_{cc}^*$&$3.96_{-0.16} ^{+0.15}$ &$-$    &$3.97$&	$3.96 \pm 0.16 $&$4.086  $&$4.95  \pm0.05$	  \\ 
		$	\Omega_{bc}^*$&$7.21  \pm0.21$   &$-$&$7.37$&$-$&$- $&$7.85  \pm0.05$		\\
		$	\Omega_{bb}^*$&$10.60 \pm0.32$        &$-$	      &$10.64$&$10.57 \pm 0.15$&$10.608	$	&$10.85  \pm0.025$ \\ 
		\hline\hline
	\end{tabular}%
	\caption{ $1P$ excited state masses (in $\mathrm{GeV}$) of doubly heavy baryons and comparison with other existing theoretical predictions.}
	\label{tab3}
\end{table}

\begin{table}[tbp]
	\centering
	\begin{tabular}{|c|c|c|c|c|}
		
		\hline Baryon& Our work &   \cite{Shah:2017liu} &  \cite{Shah:2016vmd} &    \cite{Alrebdi:2022lat} \\ \hline\hline
$	\Xi_{cc}^*$& $3.99 \pm0.16$     &$3.99$       &$-$&	$4.95  \pm0.05$  \\  
$	\Xi_{bc}^*$&$7.37  \pm0.21$                       &$7.27$            &$-$&$7.91  \pm0.05$	 \\ 
$	\Xi_{bb}^*$&$10.71  \pm0.32$  &$10.62$&$-$&$10.92  \pm0.025$ \\
$	\Omega_{cc}^*$&$4.10_{-0.17} ^{+0.16}$ &$-$    &$4.10$&$5.00  \pm0.05$	  \\ 
$	\Omega_{bc}^*$&$7.40  \pm0.21$   &$-$&$7.48$&$7.90  \pm0.05$		\\
$	\Omega_{bb}^*$&$10.82 \pm0.32$        &$-$	      &$10.74$&$10.90 \pm0.025$ \\ 
\hline\hline
\end{tabular}%
\caption{ $2S$ excited state masses (in $\mathrm{GeV}$) of doubly heavy baryons and comparison with other existing theoretical predictions.}
	\label{tab5}
\end{table}   

This study also analyzes the residues associated with the ground state, the first orbital $1P$ excitation, and the first radial $2S$ excitation  of doubly heavy baryons possessing spin-$\frac{3}{2}$, with corresponding data presented in tables~\ref{tab2},~\ref{tab4}, and~\ref{tab6}. Although earlier works using QCD sum rules have investigated the residues in the ground state \cite{Aliev:2012iv,Wang:2010vn,Bagan:1992za}, as well as the $1P$ \cite{Alrebdi:2022lat,Wang:2010it} and $2S$ excited states \cite{Alrebdi:2022lat}, our current findings offer improved accuracy for residue calculations. Although mass predictions across various approaches generally exhibit strong agreement (within reported uncertainties, with minor channel specific differences), we observe considerable variations in the reported residue values for $1S$, $1P$, and $2S$ states among different studies, despite some channel specific consistencies.
	 The inherently larger uncertainties associated with the presented residues, relative to their corresponding masses, are a consequence of their derivation: masses are obtained from a ratio of sum rules, which naturally mitigates uncertainties, whereas residues are directly extracted from a single sum rule.
	These newly presented residue values offer crucial input for future investigations into the decay channels of the doubly heavy baryons.

\begin{table}[tbp]
	\centering
	\begin{tabular}{|c|c|c|c|c|}
		\hline  Baryon& Our work &    \cite{Aliev:2012iv} &   \cite{Bagan:1992za}&  \cite{Wang:2010vn}   \\ \hline\hline
		$	\Xi_{cc}^*$&$0.15\pm0.02$                                    &$0.12\pm0.01$&$0.071 \pm 0.017 $& $0.070\pm0.017$	  \\
		$	\Xi_{bc}^*$&$0.30  \pm0.03$                                   &$0.15\pm0.01$&$-$ &$-$	 \\
		$	\Xi_{bb}^*$&	$0.50  \pm0.05$                              &$0.22  \pm0.03$&	 $0.111 \pm 0.040$&$0.161\pm0.041$ \\
		$	\Omega_{cc}^*$&$0.16  \pm0.02$                              &$0.14  \pm0.02$&$-$&$0.085\pm0.019$	 \\
		$	\Omega_{bc}^*$&$0.33 \pm0.04$                             &$0.18  \pm0.02$&$-$&	$-$ \\
		$	\Omega_{bb}^*$&$0.53\pm0.05$                        &$0.25  \pm0.03$&$-$&$0.199\pm0.048$\\
		\hline\hline
	\end{tabular}%
	\caption{Ground state $1S$ residues (in $\mathrm{GeV^3}$) of doubly heavy baryons and comparison with other existing theoretical predictions.}
	\label{tab2}
\end{table}

\begin{table}[tbp]
	\centering
	\begin{tabular}{|c|c|c|c|}
	\hline  Baryon& Our work &   \cite{Wang:2010it} &  \cite{Alrebdi:2022lat}   \\ \hline\hline
	$	\Xi_{cc}^*$&$0.17 \pm0.02$                                  &$0.095  \pm 0.020$  &$0.185 \pm 0.025 $	  \\
	$	\Xi_{bc}^*$&$0.34  \pm0.03$                                 &$-$  &$0.480 \pm 0.025$	 \\
	$	\Xi_{bb}^*$&	$0.67  _{-0.05} ^{+0.06}$                        &$0.227 \pm 0.052$      &$ 1.060 \pm 0.05$ \\
	$	\Omega_{cc}^*$&$0.20  _{-0.02} ^{+0.03}$                         &$0.116 \pm 0.022$     &$0.165 \pm 0.025$	 \\
	$	\Omega_{bc}^*$&$0.36  _{-0.04} ^{+0.05}$                      &$-$       &$0.480 \pm 0.025$ \\
	$	\Omega_{bb}^*$&$0.70\pm0.06$                  &$0.275 \pm 0.059$      &$1.075 \pm 0.05$\\
	\hline\hline
\end{tabular}%
	\caption{$1P$ excited state residues (in $\mathrm{GeV^3}$) of doubly heavy baryons and comparison with other existing theoretical predictions.}
	\label{tab4}
\end{table}

\begin{table}[tbp]
	\centering
	\begin{tabular}{|c|c|c|}
	\hline  Baryon& Our work &   \cite{Alrebdi:2022lat}   \\ \hline\hline
	$	\Xi_{cc}^*$&$0.19 \pm0.03$                                    &$0.175 \pm 0.025 $	  \\
	$	\Xi_{bc}^*$&$0.36  _{-0.03} ^{+0.04}$                                   &$0.475 \pm 0.025$	 \\
	$	\Xi_{bb}^*$&	$0.77  \pm0.07$                              &$ 1.050 \pm 0.05$ \\
	$	\Omega_{cc}^*$&$0.21 \pm0.03$                              &$0.160 \pm 0.025$	 \\
	$	\Omega_{bc}^*$&$0.38  \pm0.05$                             &$0.475\pm 0.025$ \\
	$	\Omega_{bb}^*$&$0.80\pm0.08$                        &$1.080 \pm 0.05$\\
	\hline\hline
\end{tabular}%
	\caption{$2S$ excited state residues (in $\mathrm{GeV^3}$) of doubly heavy baryons and comparison with other existing theoretical predictions.}
	\label{tab6}
\end{table}

To clarify the impact of the present extended analysis, we provide a detailed comparison with previous QCD sum rule studies, as summarized in the  tables  \ref{tab1}, \ref{tab3}, \ref{tab5}, \ref{tab2}, \ref{tab4} and \ref{tab6}. For the ground state, the masses obtained in this work are in good agreement with most existing QCD sum rule predictions. The only notable deviation appears in the $\Xi_{bc}^{*}$ channel, where our result differs from Ref. \cite{Zhang:2008rt}, while consistency is observed with the remaining studies (Refs.~\cite{Aliev:2012iv,Wang:2010vn}).
For the $1P$ excited states, our mass predictions are compatible with Ref. \cite{Wang:2010it}. However, except for the $\Omega_{bb}^{*}$ channel, noticeable numerical differences are found when compared with Ref. \cite{Alrebdi:2022lat}. 
A similar pattern is observed for the $2S$ states. Apart from the $\Omega_{bb}^{*}$ channel, our predicted masses show sizable deviations from those reported in Ref. \cite{Alrebdi:2022lat}. 

Regarding the pole residues, for the ground states our results differ from those reported in Refs. \cite{Aliev:2012iv,Wang:2010vn,Bagan:1992za} in most channels, with agreement occurring only in a limited number of cases. 
For the $1P$ states, the extracted residues also exhibit differences compared to Refs.~\cite{Wang:2010it,Alrebdi:2022lat}, except in a few specific channels. 
Furthermore, for the $2S$ states, our predicted residues differ numerically from Ref. \cite{Alrebdi:2022lat}, with the exception of the $\Xi_{cc}^{*}$  and $\Omega_{cc}^{*}$ channels where reasonable agreement is observed.

The observed differences between our results and previous QCD sum rule analyses mainly originate from several methodological aspects. 
First, different choices of the auxiliary parameters, such as the Borel window and the continuum threshold $s_{0}$, can significantly affect the extracted observables. 
Second, the use of different interpolating currents for doubly heavy baryons modifies the structure of the correlation function and the relative pole continuum contributions. 
Third, the choice of renormalization scale $\mu$ plays an essential role in heavy–quark systems. In the present work, following standard QCD sum rule prescriptions, we vary the renormalization scale in the range $\mu = (2\text{--}4)~\mathrm{GeV}$. The final reported masses and residues correspond to the averaged values obtained within this window, and the associated scale dependence is included in the quoted uncertainties. Moreover, our operator product expansion is carried out consistently up to dimension ten condensates. In many previous QCD sum rule studies of these systems, the expansion was truncated at lower dimensions (typically dimension five or dimension six), which can partially explain the numerical differences observed here. In order to estimate the uncertainties associated with the factorization (vacuum saturation) approximation for higher-dimensional condensates (dimension $\geq 6$), we vary the parameter $\kappa$ in the range $\kappa = (1\text{--}5)$. The maximal variation induced by this procedure is incorporated into the total uncertainties of both the masses and the residues. Numerically, this effect contributes less than approximately $2\%$ to the total uncertainties.

\section{conclusion}\label{Sum} 

The existence of baryons composed of two heavy quarks presented a long standing challenge in the quark model, as experimental observation remained elusive. The discovery of the doubly charmed baryon $	\Xi_{cc}$ by the LHCb Collaboration has spurred significant interest in doubly heavy baryons. Despite a discrepancy in its measured mass compared to earlier SELEX Collaboration findings, this discovery fuels optimism for the identification of other doubly heavy charmed and bottom baryons. Continued experimental efforts are anticipated to reveal more of the doubly heavy baryon spectrum predicted by quark models. Consequently, a thorough examination of their properties, particularly spectroscopic parameters, is highly valuable. This work focuses on investigating the spectroscopic characteristics of doubly heavy spin-$\frac {3}{2} $ baryons, with the aim of providing insights to aid experimental groups in analyzing forthcoming data.

Accurate spectroscopic parameters, including masses and residues, are essential for investigating the interactions and decays of doubly heavy baryons. Theoretical calculations of these parameters also serve as crucial guides for experimental searches for yet-undiscovered baryon states predicted by quark models. This study focuses on precisely determining the spectroscopic parameters of doubly heavy baryons across the ground state ($1S$), first orbital excitation ($1P$), and the first radial excitation ($2S$). Using the QCD sum rule approach, a reliable and predictive technique based on the QCD Lagrangian, we include nonperturbative operators up to dimension ten to enhance accuracy. Following the determination of auxiliary parameters, we evaluate the mass spectra and residue for these specific baryon states, presenting them in table~\ref{tab:results} and comparing our findings with results from other theoretical frameworks.
The outcomes of this study are poised to aid experimental groups in their quest to discover the yet-unseen members of the doubly heavy baryon family. Additionally, these predictions offer a valuable basis for comparison with future theoretical work employing different nonperturbative methods. The importance of residue values, critical for computing any physical quantities related to baryon interactions and decays, cannot be overstated. Our  results provide the necessary input data for comprehensive studies of the decay characteristics of the considered doubly heavy baryons in their ground and excited configurations. These investigations are essential for determining the decay widths and estimating the lifetimes of these baryons.

\section*{ACKNOWLEDGMENTS}
Financial support for this work was provided in part by the Iran National Science Foundation (INSF) under Grant No. 4037888. Additionally, K. Azizi thanks the CERN Theory Department for their support and hospitality.

\section*{APPENDIX: SPECTRAL DENSITIES UTILIZED IN OUR CALCULATIONS}
This appendix details the specific expressions for each component of the spectral density as obtained through our calculations.

 \begin{align}
 	\rho^{(pert)}_{~\not\!q g_{\mu\nu}}(s)&=\frac{1}{ 32 \pi^4 }	\int_{0}^{1} \, du \int_{0}^{1-u}
  dv \, \frac{D_1   \Theta(D_1)}{ Z_1^4 Z_3^2} \nonumber\\
  &\Bigg\{ 3  Z_1^2  D_1 u v Z_2    
 + \Bigg(Z_3 \Big( m_Q u Z_1^2 (m_{Q'} v - m_q Z_2) - v Z_2 (m_{Q'} m_q Z_1^2 - 3 s u^2 v Z_2) \Big) \Bigg) \Bigg\},
 \end{align}
 
  \begin{align}
 	\rho^{(pert)}_{ g_{\mu\nu}}(s)&=\frac{1}{ 32 \pi^4 }	\int_{0}^{1} \, du \int_{0}^{1-u}
 	dv \, \frac{D_1   \Theta(D_1)}{ Z_1^3 Z_3^2} \nonumber\\
 	&\Bigg\{ \Bigg(Z_1^3 D_1 (m_Q u + m_{Q'} v - m_q Z_2)  \Bigg)  
 	+ \Bigg(Z_3  \Big(-s u v Z_2 (m_{Q'} v - m_q Z_2) + m_{Q} (3 m_{Q'} m_q Z_1^2 - s u^2 v Z_2)\Big) \Bigg) \Bigg\},
 \end{align}

\
 \begin{align}
	\rho^{(3d)}_{~\not\!q g_{\mu\nu}}(s)&=\frac{1}{24 \pi^2}	\int_{0}^{1} \, du \int_{0}^{1-u}
	dv \,    \,  \langle \overline{q} q \rangle \,  \Theta(D_2) \Bigg\{ -m_{Q} u - (m_{Q'} - 3 m_q u) Z_4 \Bigg\},&&&&&&&&&
\end{align}

\begin{align}
	\rho^{(3d)}_{ g_{\mu\nu}}(s)&=\frac{1}{24 \pi^2}	\int_{0}^{1} \, du \int_{0}^{1-u}
	dv \,    \,  \langle \overline{q} q \rangle \,  \Theta(D_2) \Bigg\{ -2 D_2 + m_{Q} (-3 m_{Q'} + m_q u) - (-m_{Q'} m_q + s u) Z_4 \Bigg\},&&&&&&&&&
	\end{align}

 \begin{align}
	\rho^{(4d)}_{~\not\!q g_{\mu\nu}}(s)&=\frac{1}{3072  \pi^2}	\int_{0}^{1}  \, du \int_{0}^{1-u}
	dv \,  \Big\langle\frac{\alpha_{s}GG}{\pi}\Big\rangle\ \,  \frac {u v Z_2 (16 u^2 - 16 u Z_3 - 15 v Z_3)  \,  \Theta(D_3) } {Z_1^4},&&&&&&&&&&&&&&&&&&&&&&&&&&&&&&&&&&&&&&&&&&&&&&&&&&&&&
\end{align}

\begin{align}
	\rho^{(4d)}_{ g_{\mu\nu}}(s)&=\frac{-1}{1152  \pi^2}	\int_{0}^{1}  \, du \int_{0}^{1-u}
	dv \,  \Big\langle\frac{\alpha_{s}GG}{\pi}\Big\rangle\ \,  \frac {  \,  \Theta(D_3) } {Z_1^4} \Bigg\{ 2 m_{Q'} - 13 (m_{Q} + m_{Q'} + m_q) u + \Big(14 m_{Q} + 11 m_{Q'}  \nonumber\\
	&+ 
	13 m_q\Big) u^2 v^3+ \Big(13 (m_{Q} + m_q) u + 3 m_{Q'} (-8 + 3 u)\Big) v^4 + 
	20 m_{Q'} v^5 - 4 m_{Q} u^3 (-1 + 5 u) Z_4  \nonumber\\
	&- 
	u^2 v Z_4 \Big(m_{Q} (-17 + 11 u) - 13 (m_{Q'} + m_q) Z_4\Big) + 
	u v^2 m_{Q} \Big(13 + 3 u (-15 + 8 u)\Big) - \Big(-17 m_{Q'}  \nonumber\\
	&- 13 m_q + 
	26 (m_{Q'} + m_q) u\Big) Z_4\Bigg\},&&&&&&&&&&&&&&&&&&&&&&&&&&&&&&&&&&&&&&&&&&&&&&&&&&&&&
	\end{align}

 \begin{align}
	\rho^{(5d)}_{~\not\!q g_{\mu\nu}}(s)=0,&&&&&&&&&&&&&&&&&&&&&&&&&&&&&&&&&&&&&&&&&&&&&&&&&&&&&&&&&&&&&&&&&&&&&&&&&&&&&&&&&&&&&&&&&&&&&&&&
\end{align}

\begin{align}
	\rho^{(5d)}_{g_{\mu\nu}}(s)&=\frac{5}{96 \pi^2}	\int_{0}^{1} \, du \int_{0}^{1-u}
	dv \,   \, \langle \overline{q}g_s\sigma Gq\rangle \, u \, Z_4   \,  \Theta(D_4)  ,&&&&&&&&&&&&&&&&&&&&&&&&&&&&&&&&&&&&&&&&&&&&&&&&&&&&&&&&&&&&
\end{align}

where

\begin{eqnarray}
		D_1&=&-\frac{Z_3}{(u^2 - (u + v) Z_3)^2} \Bigg(s u v Z_2 + m_{Q'}^2 u (u^2 - (u + v) Z_3) + m_Q^2 v \Big(u^2 - (u + v) Z_3\Big) \Bigg),\nonumber\\
		D_2&=&-m_{Q'}^2 Z_4 - u (m_{Q}^2 - s Z_4),\nonumber\\
		D_3&=&\frac{1}{Z_1^2} \Bigg(\Big(m_{Q}^2 u Z_1 + v (m_{Q'}^2 Z_1 + s u Z_2)\Big) Z_3 \Bigg) ,\nonumber\\
		D_4&=&-m_{Q'}^2 Z_4 - u^2 (m_{Q}^2 - s Z_4),
\end{eqnarray}

and we have also described

\begin{eqnarray}
Z_1&=& u^2 + u (-1 + v) + (-1 + v) \, v,\nonumber\\
Z_2&=&1 - u - v,\nonumber\\
Z_3&=&1 - v,\nonumber\\
Z_4&=&1 -u.
\end{eqnarray}


\end{document}